\documentclass[a4paper,12pt]{article}
\usepackage{subfigure}
\usepackage{graphicx}
\usepackage{epsfig}
\usepackage[psamsfonts]{amssymb}
\newcommand{\No}{N\b{\tiny o}}
\newcommand{\F}{\ensuremath{\Phi}}
\begin{document}
\thispagestyle{headings}
\title{Measurement of the $\phi$ meson parameters with CMD-2 detector
at VEPP-2M collider
\thanks{Work is supported in part by grants RFBR-98-02-17851, 
RFBR-99-02-17053, RFBR-99-02-17119 }}

\author{
R.R.Akhmetshin,  E.V.Anashkin, M.Arpagaus,  V.M.Aulchenko, \and
V.Sh.Banzarov,  L.M.Barkov,  S.E.Baru,  N.S.Bashtovoy,  A.E.Bondar,   \and
D.V.Chernyak,  A.G.Chertovskikh,  A.S.Dvoretsky, S.I.Eidelman,    \and
G.V.Fedotovich,  N.I.Gabyshev,  A.A.Grebeniuk, D.N.Grigoriev,   \and
B.I.Khazin, I.A.Koop, P.P.Krokovny,  L.M.Kurdadze,  A.S.Kuzmin, \and
I.B.Logashenko, P.A.Lukin\thanks{contact person. e-mail:
P.A.Lukin@inp.nsk.su}, A.P.Lysenko, K.Yu.Mikhailov, \and
I.N.Nesterenko, V.S.Okhapkin, E.A.Perevedentsev, A.A.Polunin, \and
E.G.Pozdeev, V.I.Ptitzyn, T.A.Purlatz,  N.I.Root,  \and 
A.A.Ruban, N.M.Ryskulov, A.G.Shamov,  Yu.M.Shatunov,  \and 
A.I.Shekhtman, B.A.Shwartz, V.A.Sidorov, A.N.Skrinsky, \and 
V.P.Smakhtin, I.G.Snopkov, E.P.Solodov, P.Yu.Stepanov, \and  
A.I.Sukhanov, Yu.V.Yudin,  S.G.Zverev \\ \vspace{2mm}
{\it Budker Institute of Nuclear Physics, Novosibirsk, 630090,
Russia} \\ \vspace{2mm}
J.A.Thompson \\ \vspace{2mm}
{\it University of Pittsburgh, Pittsburgh, PA, 15260, USA}
}

\date{}

\maketitle

\begin{abstract}
About 300 000 \(
e^{+}e^{-}\rightarrow \phi\rightarrow K^0_L K^0_S \) events 
in the center of mass energy range from 984 to 1040 MeV were
used for the measurement of the $\phi$ meson parameters.
The following results have been obtained: 
$\sigma_0 = (1367 \pm 15 \pm 21)$ nb,  
$m_{\phi}=(1019.504\pm 0.011\pm 0.033)$ MeV/c$^2$, 
$\Gamma_\phi=(4.477\pm 0.036\pm 0.022)$ MeV,
$\Gamma_{e^+e^-}\cdot$B$(\phi\to K^0_L K^0_S) = (4.364 \pm 0.048 \pm
0.065)\cdot 10^{-4}$ MeV.
\end{abstract}
\section{Introduction}
%\par Precise measurements of $\phi$ meson parameters provide valuable
%information for physics of light quarks. The decay 
%$\phi\rightarrow K^0_L K^0_S$ is one of the
%dominant of the $\phi$ meson and can be used for $\phi$ parameter
%measurement. These measurements were performed by various groups
%\cite{ZPHY,PR,SJNP}. 
%\par Since 1992 the CMD-2 detector \cite{CMD-2.1,CMD-2.2} has been
%running at the high luminosity collider VEPP-2M \cite{VEPP}. The first
%results from the CMD-2 detector were published in \cite{BW} with
%relatively small integrated luminosity about 300 nb$^{-1}$ on
%study of the four major modes of the $\phi$ meson decay.
%\par This paper presents the results of the $\phi$
%meson parameters measurement in the $\phi\rightarrow K^0_L K^0_S$ 
%decay channel. 
%\par The data discussed in present work were collected in four scans of
%$2E_{beam}$ = 984 -- 1040 MeV energy range. One scan was performed
%in 1994 with energy measurement at each energy point by
%resonance depolarization method
%\cite{Dep} and three scans were in 1996 with total integrated luminosity
%0.18 pb$^{-1}$ and 2.19 pb$^{-1}$ respectively. The number of the
%$\phi$ meson decays is $3.4\times10^6$.
%\par The decay mode $\phi\to K^0_L K^0_S$ is one of the dominant for
%the $\phi$ meson and can be used for the determination of such basic
%properties of the resonance as its mass, total and leptonic width.
\par This paper presents a precise determination of the mass,
and total and leptonic widths of the $\phi$, based on one
of its dominant decay modes, $\phi\to K^0_L K^0_S$. High precision 
measurements of the $\phi$ meson parameters provide
valuable information for various theoretical models describing
interactions of light quarks.
%\par In this work we report on the high precision measurement of the
%$\phi$ meson parameters performed with the CMD-2 detector 
%\cite{CMD-2.1,CMD-2.2} which has been running at the high luminosity
%collider VEPP-2M \cite{VEPP} since 1992.
\par The results of this work are based on data collected with the 
CMD-2 detector \cite{CMD-2.1,CMD-2.2} which has been running at the 
high luminosity collider VEPP-2M \cite{VEPP} since 1992.
The data come from four scans of the center of mass energy range 
2$E_{beam}$ from 984 to 1040 MeV. In the first scan, performed in 1994 
with integrated luminosity of 0.18 pb$^{-1}$, the resonance depolarization
method \cite{Dep} was used for precise beam energy calibration at each
point. Three other scans, performed in 1996, corresponding to integrated 
luminosity of 2.19 pb$^{-1}$, do not have resonance depolarization 
information. The collected integrated 
luminosity of 2.37 pb$^{-1}$ corresponds to 3.4$\times 10^6$ $ \phi$
meson decays.
\section{CMD-2 detector}
\par The CMD-2 detector has been described in detail elsewhere 
\cite{CMD-2.1,CMD-2.2}. It is a general purpose detector consisting of
a drift chamber (DC) with about 250 $\mu$ resolution in the plane
transverse to the
beam and proportional Z-chamber (ZC) used for the trigger, both inside a
thin (0.38 $X_0$) superconducting solenoid with a field of 1 T.
\par The barrel calorimeter placed outside the solenoid consists of
892 CsI crystals of 6$\times$6$\times$15 cm$^3$ size. It covers polar
angles from 0.8 to 2.3 radian. The energy resolution for the photons
in the CsI calorimeter is about 8 \% in the energy range from 100 to
700 MeV.
% BGO endcap calorimeter consists of 680
% 25$\times$25$\times$150 mm$^3$ crystals and covers polar angles from
% 0.28 to 0.86 and from 2.28 to 2.86 radians for first and second endcap
% respectively.
% \par The muon system is composed of two layers of streamer tubes 
% separated by the 15 cm thick iron magnet yoke. Its spatial resolution
% is about 2.5 cm.
\par The trigger signal is generated either by the charged trigger based
on DC and Z-chamber hits \cite{TF} or by the neutral trigger \cite{NT}
which takes into account the number of clusters detected in the CsI 
calorimeter as well as the total energy deposition.
Two independent triggers of CMD-2 can be used to study the trigger 
efficiency.
%The data discussed in present work were collected during 1994 physics
%runs at $\phi$ meson energy region with resonant depolarization method
%energy measurement at each point and during three scans of $2E_{beam}$
%= 984 -- 1040 MeV energy range in 1996. Total integrated luminosity is
%about 2.37 pb$^{-1}$, corresponding to $3.4\times 10^6$ of $\phi$
%mesons.
% The data 
% collected during the 1994 and 1996 physics runs  
% have been  analysed with the luminosity integral  2.37 pb$^{-1}$
% in $2E_{beam}$ = 984 -- 1040 MeV energy range, corresponding
% to $3.4\times 10^6$ of $\phi$ mesons.
\section{Analysis} 
\par
\begin{figure*}
\epsfig{file=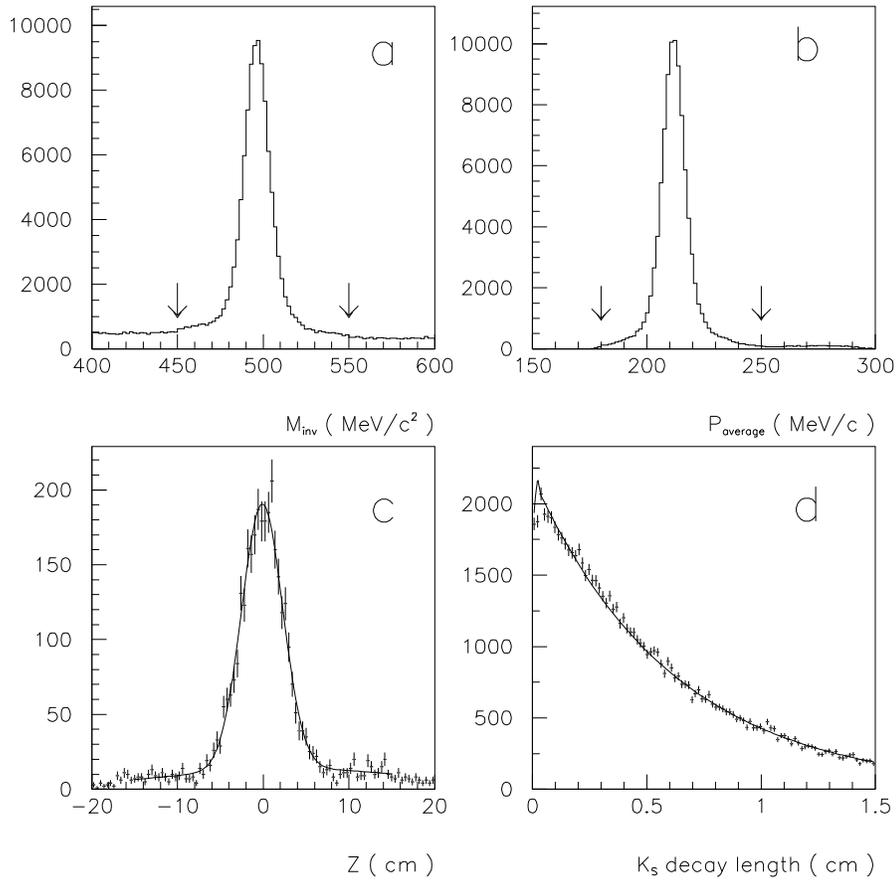,width=0.95\textwidth}
\caption{Distributions used for the selection of $e^+ e^- \to \phi
\to K^0_L K^0_S$ events. a - Invariant mass of two tracks; 
b - Average momentum of two tracks;
c - Z-coordinate of the vertex; d - Decay length of $K^0_S$ meson. }
\label{fig:cuts}
\end{figure*}

 The process $\phi\rightarrow K^0_L K^0_S$ was detected using
the $K^0_S$ decay into two charged pions.
Events  were selected according to the following conditions:
\begin{itemize}
\item Two oppositely charged tracks were found coming from the vertex closest
to the beam.  
\item Both tracks had polar angles $0.95 < \theta_{1,2} < \pi - 0.95 $. 
The distance between the beam and the vertex in the $R-\varphi$ plane is:
$R_{vert} < 1.5$  cm.
\item The invariant mass of the two vertex tracks, taken as pions,  was 
$450 < M_{inv} < 550$ MeV/c$^2$ (see Fig.~\ref{fig:cuts}a)
and the missing momentum satisfied the conditions
$60 < P_{mis} < (\sqrt{E_{beam}^2 - m_{K^0}^2} + 40) $ MeV/c,  
where $m_{K^0} = 497.672$  MeV/c$^2$  is the neutral kaon mass \cite{pdg}.
\item Track momenta were
$ 140 < P_{1,2} < 300 $ MeV/c
and  the average momentum was
$180 < (P_1 + P_2)/2 < 250 $ MeV/c as shown in Fig.~\ref{fig:cuts}b. 
\end{itemize}
\par At each energy the cross section
of $K^0_L K^0_S$ production
was calculated according to the formula:
\begin{eqnarray*}
\sigma(e^+ e^-\rightarrow K^0_L K^0_S) = 
\frac{N}
{\varepsilon\cdot L\cdot(1+\delta_{rad})\cdot(1+\delta_{loss})
\cdot B(K^0_S\to\pi^+\pi^-)},
\end{eqnarray*}
where $N$ is the number of events; $L$ is the integrated luminosity 
determined from large angle Bhabha events
with the help of the procedure described in \cite{Prep99}; 
$\delta_{rad}$ is the radiative correction  calculated according to  
\cite{Rad}; $\delta_{loss}$ is a correction  for event losses due to
decays in flight and nuclear interactions of the charged pions; 
B$(K^0_S\to\pi^+\pi^-) = 0.6861 \pm 0.0028 $
is the branching ratio of the decay $K^0_S\to\pi^+\pi^-$ from \cite{pdg};
and $\varepsilon$ is a 
product of the reconstruction
efficiency and geometrical efficiency (acceptance),
$\varepsilon = \varepsilon_{rec}\varepsilon_{geom}$.
The acceptance is the probability to detect two pions from the $K^0_S$
decay within a solid angle determined by the cuts above.
\par To obtain the number of $K^0_L K^0_S$ events,
cosmic ray events were removed by subtracting a smooth background 
in the distribution over the Z-coordinate of the vertex as demonstrated 
in Fig.~\ref{fig:cuts}c. 
A sum of two gaussian functions was used as a fitting curve, one
gaussian describing $K^0_L K^0_S$ events and the other describing the
shape of background. To estimate a systematic uncertainty because of the
background subtraction, we used an alternative assumption that the
background was described by a straight line with a slope. The difference
in the number of events obtained by two methods was 0.3\%.      
The distribution over the $K^0_S$ meson decay 
length after all selections 
and background subtraction is shown in Fig.~\ref{fig:cuts}d. The
exponential curve in the Figure was calculated using the generally accepted
value of the $K^0_S$ lifetime \cite{pdg} and agrees well with the data.  
\par The background from the $\phi\to K^+ K^-$ and 
$\phi\to \pi^+ \pi^- \pi^0$
decays was estimated from Monte Carlo simulation to be less than 0.04\%.
%\par With the above cuts   $2.97\times 10^5$ of $K^0_L K^0_S$ events
%have been selected. The distribution over the solid angle between tracks
%demonstrated in Fig \ref{Psi} confirms the hypotesis that the tracks
%are pions from the $K^0_S$ decay. The Fig. \ref{ThetaKL} shows
%distribution over the $K^0_L$ polar angle calculated with the help
%of the tracks momenta and angles information. The shape of the
%distribution is in agreement with the expected one for the $K \bar K$.
\par
\begin{figure*}
\includegraphics[width=0.48\textwidth]{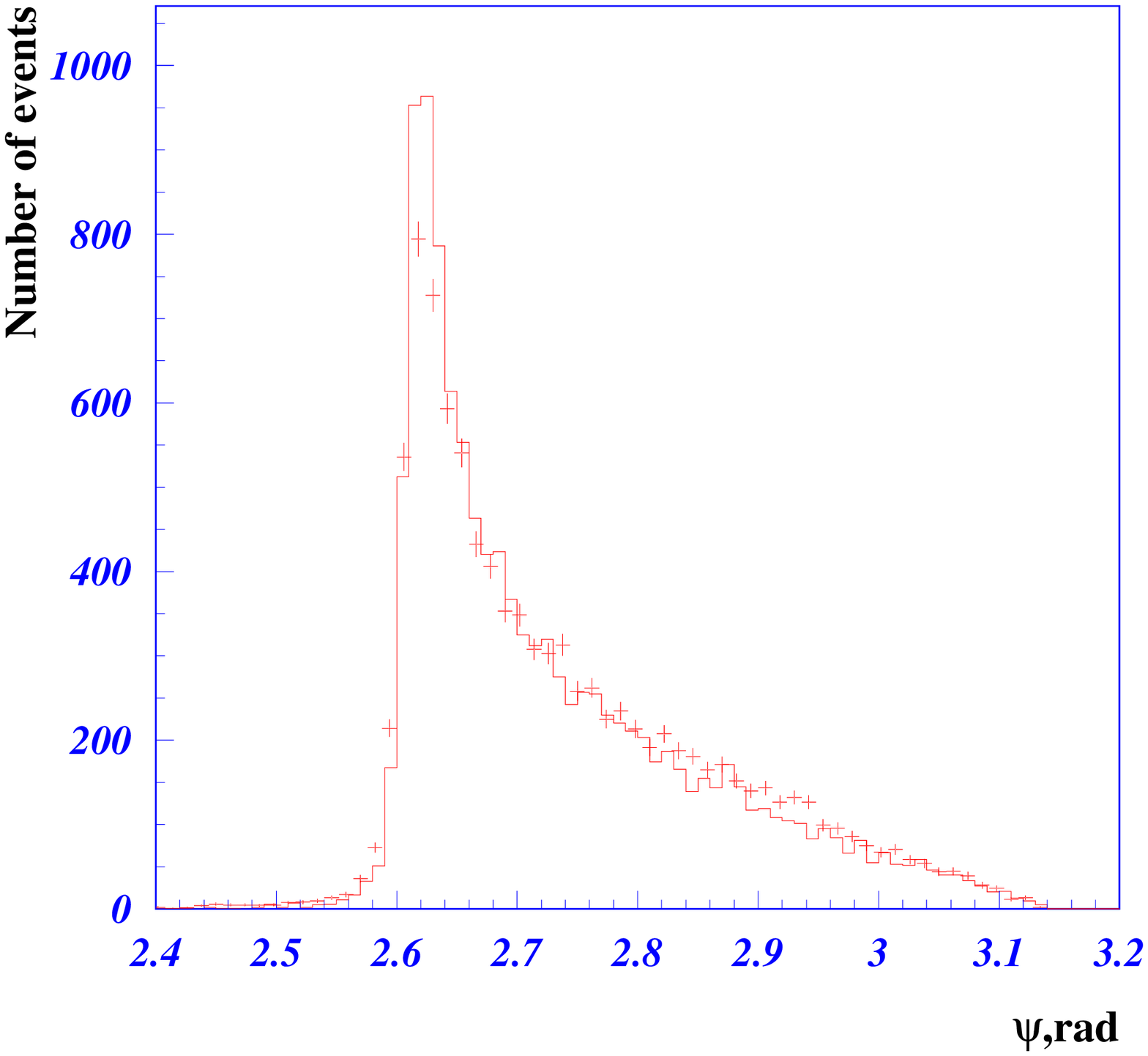}
\hfill
\includegraphics[width=0.48\textwidth]{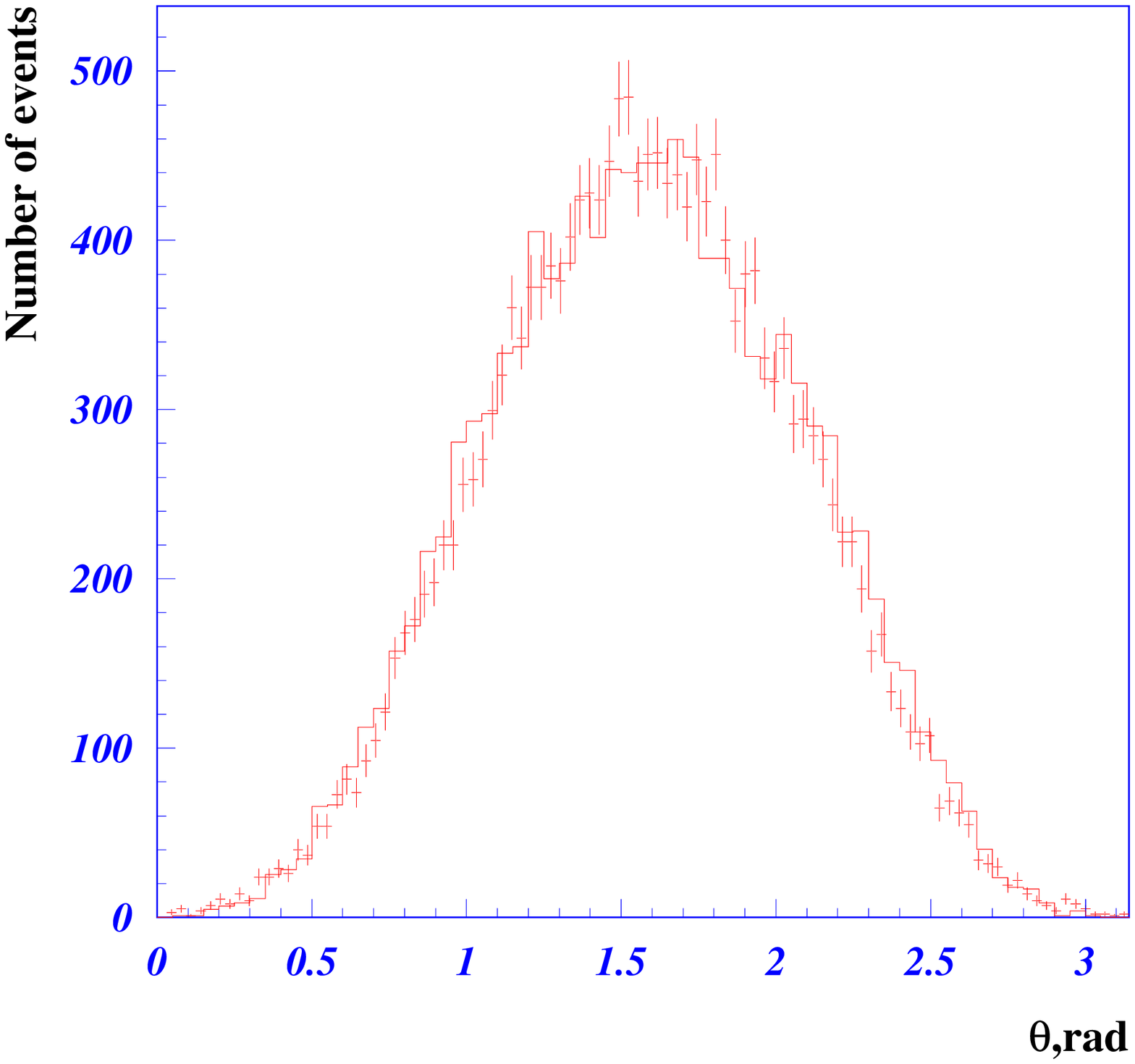}
\\
\parbox[t]{0.48\textwidth}{\caption{  Distribution of the space
angle between two tracks. The histogram is simulation, points with errors are
data.} \label{Psi}}
\hfill
\parbox[t]{0.48\textwidth}
{\caption{Distribution of the $K^0_L$ polar angle. The histogram is
simulation, points with errors are data. } \label{ThetaKL}}
\end{figure*}

 With the above cuts   $2.97\times 10^5$ $K^0_L K^0_S$ events
have been selected. The distribution of the space angle between 
the tracks presented in Fig.~\ref{Psi} shows a minimal angle 
typical for pions from the $K^0_S$ decay. Figure~\ref{ThetaKL} shows
the distribution for the $K^0_L$ polar angle calculated from  the 
momenta of the tracks as well as their angles. The shape of the
distribution is consistent with that expected for the $K \bar K$.
%\par About 50 \% of $K^0_L$ have a cluster in the CsI calorimeter because of
%the nuclear interaction. 
%Using the polar and azimuthal angles of this cluster as well as the angles of
%the clusters produced by pions from the decay $K^0_S\rightarrow
%\pi^+\pi^-$, one can reconstruct the $\phi\to K^0_L K^0_S$ event 
%without DC information.  
% Adding two clusters from pions from the $K^0_S$ decay
% one can reconstruct the $\phi\to K^0_L K^0_S$ event without DC information. 
%These ``test'' events were used to determine the charged trigger efficiency. 
%\par To study the 
%reconstruction efficiency, the same events have been used with an additional
%requirement of ZC hits and at least one track in DC to reduce background
%from $\phi\to \pi^+ \pi^- \pi^0$
%decays was estimated from Monte Carlo simulation to be less than 0.04\%.
%\par With the above cuts   $2.97\times 10^5$ of $K^0_L K^0_S$ events
% have been selected.
\par About 50 \% of the $K^0_L$ events have a cluster in the CsI 
calorimeter resulting from a nuclear interaction. 
Using the polar and azimuthal angles of this cluster as well as the angles of
the clusters produced by pions from the decay $K^0_S\rightarrow
\pi^+\pi^-$, one can reconstruct the $\phi\to K^0_L K^0_S$ event 
without DC information.  
% Adding two clusters from pions from the $K^0_S$ decay
% one can reconstruct the $\phi\to K^0_L K^0_S$ event without DC information. 
Such  ``test'' events were used to determine the charged trigger efficiency. 
\par To study the 
reconstruction efficiency, the same events have been used with an additional
requirement of ZC hits and at least one track in DC to reduce background
from the $\phi\to  \pi^+ \pi^- \pi^0$ decays. 
\par About 9 000 ``test'' events were selected.
%
% These ``test'' events were used to determine reconstruction efficiency.
%To reduce the background from $\phi\to 3\pi$ decays
%ZC hits and presence at least one track in DC were required.
% \par 
% Two independed triggers of CMD-2 (charged and neutral) can be used
% to study trigger inefficiency. 
% Absence of a charged trigger signal for 
% the ``test'' events without a requirement of a reconstructed track in DC 
% provides the charged trigger inefficiency.
The detailed description of the efficiency determination procedures 
can be found in \cite{phys54}.
\par The acceptance as well as the correction for decays in flight was
calculated by  Monte Carlo simulation.
\par Typical values of the efficiencies and 
corrections 
% for decays in flight and nuclear interaction  
are presented in Table
\ref{tab:eff} for  $2E_{beam} = 1020.0$ MeV 
(the $\phi$ meson peak).
\par For the 1994 data the beam energy at each point was measured by
the resonance 
depolarization method \cite{Dep}. For the 1996 data the beam 
energy was determined from the momenta of charged
kaons and  the analysis of the magnetic field of the collider \cite{Prep99}.
\begin{table*}
\caption{Efficiencies, corrections and their errors at 
$2E_{beam} = 1020.0$ MeV}
\vspace*{5mm}
\begin{tabular}{|c|c|c|c|}
\hline
Efficiency            & Value,\% & Stat. error,\%  & Syst. error, \% \\
\hline
$\varepsilon_{rec}$   & 90.30    &  1.51           &  0.70             \\
\hline
$\varepsilon_{trig}$  & 98.32    &  0.93           &  0.50             \\ 
\hline
$\varepsilon_{geom}$  & 46.87    &  0.35          &  0.90            \\
\hline
1+$\delta_{loss}$    & 95.53    &  0.17           &  0.10       \\
\hline
1+$\delta_{rad}$     & 77.67    &  0.26           &  0.50       \\
\hline
\end{tabular}
%\caption{Efficiencies, corrections and their errors at 
%$2E_{beam} = 1020.0$ MeV} 
\label{tab:eff}
\end{table*}
\par
Tables \ref{Data94},\ref{Data96} show the number of selected events, 
integrated luminosity and measured cross section at each energy. 

\begin{table*}
\caption{Energy, number of events, integrated luminosity and measured 
cross section for the 1994 data. }
\vspace*{5mm}
\begin{tabular}{|c|c|c|c|c|}
\hline
\No & $E_{c.m.}$, MeV & $N_{K_L K_S}$ & L, nb$^{-1}$ & $\sigma$, nb \\ 
\hline 
  1 &$1010.272\pm 0.030$ &  135$\pm$ 13 & 16.98$\pm$0.25 &   43.07$\pm$2.84  \\    
  2 &$1017.086\pm 0.020$ & 8487$\pm$ 95 & 80.50$\pm$0.80 & 626.73$\pm$27.92 \\   
  3 &$1018.136\pm 0.018$ & 4697$\pm$ 78 & 27.68$\pm$0.36 & 921.22$\pm$81.49 \\   
  4 &$1018.956\pm 0.018$ & 5347$\pm$ 74 & 25.59$\pm$0.34 &1327.92$\pm$58.83 \\   
  5 &$1019.214\pm 0.020$ & 7610$\pm$ 89 & 27.23$\pm$1.23 &1325.12$\pm$92.42 \\   
  6 &$1019.986\pm 0.020$ & 9372$\pm$103 & 35.44$\pm$0.39 &1341.46$\pm$52.03 \\   
  7 &$1020.128\pm 0.020$ & 3725$\pm$ 62 & 16.00$\pm$0.20 &1164.35$\pm$91.97 \\
  8 &$1021.850\pm 0.022$ & 1900$\pm$ 46 & 17.34$\pm$0.27 & 615.33$\pm$55.78 \\   
  9 &$1023.972\pm 0.020$ & 2645$\pm$ 90 & 48.16$\pm$0.45 & 277.42$\pm$48.23 \\
\hline
\end{tabular}
% \caption{Energy, number of events, integrated luminosity and measured 
% cross section for the 1994 data. } 
\label{Data94}
\end{table*}   
\begin{table*}
\caption{Energy, number of events, integrated luminosity and measured 
cross section for the 1996 data. }
\vspace*{5mm}
\begin{tabular}{|c|c|c|c|c|}
\hline
\No & $E_{c.m.}$, MeV & $N_{K_L K_S}$ & L, nb$^{-1}$ & $\sigma$, nb \\
\hline
\multicolumn{5}{|c|}{Scan-2}\\
\hline
  1 &1004.25$\pm$0.36 &   77$\pm$ 37 &  31.23$\pm$0.30 &   11.83$\pm$ 4.26  \\  
  2 &1010.53$\pm$0.11 &  216$\pm$ 19 &  21.62$\pm$0.27 &   49.24$\pm$3.84  \\  
  3 &1016.36$\pm$0.04 & 2265$\pm$ 50 & 29.13$\pm$0.33  &  398.06$\pm$12.26  \\  
  4 &1017.19$\pm$0.03 & 6552$\pm$ 86 & 55.72$\pm$0.40  &  603.50$\pm$12.11  \\  
  5 &1018.06$\pm$0.03 & 8214$\pm$ 94 & 47.04$\pm$0.36  &  903.80$\pm$22.06  \\  
  6 &1018.99$\pm$0.02 &15755$\pm$127 & 63.79$\pm$0.42  & 1295.37$\pm$22.60  \\  
  7 &1020.00$\pm$0.02 &18403$\pm$137 & 71.46$\pm$0.45  & 1228.57$\pm$18.92  \\  
  8 &1020.96$\pm$0.02 &10855$\pm$108 & 52.13$\pm$0.38  &  926.26$\pm$19.72  \\  
  9 &1021.88$\pm$0.05 & 2800$\pm$ 55 & 19.71$\pm$0.24  &  617.27$\pm$26.36  \\  
 10 &1027.69$\pm$0.09 &  862$\pm$ 70 & 21.33$\pm$0.29  &  136.53$\pm$14.05  \\  
 11 &1033.63$\pm$0.36 &  492$\pm$ 84 & 28.76$\pm$0.29  &   52.61$\pm$14.17  \\  
 12 &1039.48$\pm$0.36 &  285$\pm$ 19 & 22.89$\pm$0.26  &   34.86$\pm$4.98  \\
\hline
\multicolumn{5}{|c|}{Scan-3}\\
\hline    
  1 &1004.64$\pm$0.36&   39$\pm$ 28 &  42.60$\pm$0.45 &   4.55$\pm$ 2.55 \\   
  2 &1011.30$\pm$0.06&  640$\pm$ 45 &  55.15$\pm$0.90 &  62.60$\pm$ 3.62 \\   
  3 &1015.99$\pm$0.03& 4570$\pm$ 86 &  61.59$\pm$0.41 & 378.37$\pm$ 7.97 \\   
  4 &1016.93$\pm$0.02& 7541$\pm$106 &  67.59$\pm$0.43 & 573.30$\pm$10.29 \\   
  5 &1017.91$\pm$0.02&17836$\pm$164 & 100.02$\pm$0.52 & 924.53$\pm$11.52 \\   
  6 &1019.04$\pm$0.02&27070$\pm$190 & 103.52$\pm$0.53 &1345.42$\pm$12.95 \\   
  7 &1019.95$\pm$0.02&25681$\pm$183 &  98.95$\pm$0.52 &1296.68$\pm$14.12 \\  
  8 &1020.86$\pm$0.02&18433$\pm$158 &  91.79$\pm$0.51 & 950.55$\pm$12.52 \\   
  9 &1021.74$\pm$0.03& 8685$\pm$110 &  58.36$\pm$0.41 & 664.54$\pm$12.07 \\   
 10 &1022.67$\pm$0.05& 3326$\pm$ 69 &  39.55$\pm$0.35 & 458.51$\pm$20.41 \\   
 11 &1028.36$\pm$0.09&  845$\pm$ 44 &  25.13$\pm$0.27 & 110.09$\pm$ 8.24 \\   
 12 &1034.06$\pm$0.36&  487$\pm$ 41 &  28.44$\pm$0.28 &  50.00$\pm$ 8.38 \\   
\hline
\multicolumn{5}{|c|}{Scan-4}\\
\hline  
  1 &1004.19$\pm$0.36 &   46$\pm$ 10 &   25.08$\pm$0.26 &   11.25$\pm$ 2.36 \\   
  2 &1011.30$\pm$0.07 &  382$\pm$ 37 &   32.12$\pm$0.30 &   66.11$\pm$ 9.68 \\   
  3 &1015.91$\pm$0.03 & 2478$\pm$ 61 &   33.16$\pm$0.30 &  391.00$\pm$18.90 \\   
  4 &1016.94$\pm$0.02 & 6735$\pm$ 95 &   61.84$\pm$0.41 &  600.76$\pm$18.22 \\   
  5 &1017.92$\pm$0.02 &14703$\pm$140 &   79.37$\pm$0.47 &  989.72$\pm$18.87 \\   
  6 &1018.76$\pm$0.02 &21322$\pm$163 &   84.61$\pm$0.48 & 1287.46$\pm$19.00 \\   
  7 &1019.68$\pm$0.02 &22823$\pm$167 &   86.89$\pm$0.49 & 1356.42$\pm$20.34 \\   
  8 &1020.68$\pm$0.03 & 8800$\pm$105 &   40.75$\pm$0.34 & 1058.66$\pm$29.39 \\   
  9 &1021.60$\pm$0.03 & 6237$\pm$ 91 &   41.57$\pm$0.34 &  699.48$\pm$24.72 \\   
 10 &1022.59$\pm$0.04 & 4072$\pm$ 74 &   38.15$\pm$0.32 &  465.68$\pm$19.25 \\   
 11 &1028.41$\pm$0.07 &  961$\pm$ 45 &   30.77$\pm$0.31 &  104.97$\pm$12.18 \\    
\hline
\end{tabular}
%\caption{Energy, number of events, integrated Luminosity and measured 
% cross section for the 1996 data. } 
\label{Data96}
\end{table*}
\section{Data fits  and  {\boldmath $\phi$} meson 
parameters}
\par  The experimental points were  
fit with a function  which includes the contributions
of the $\rho$, $\omega$, $\phi$   as well as higher
resonances $\omega(1420)$, $\rho(1450)$, $\omega(1600)$, $\phi(1680)$,
$\rho(1700)$ (below referred to as $\omega', \rho', \omega'', 
\phi', \rho''$ respectively) and the nonresonant background:
\begin{eqnarray*}
\sigma_K (s) = \frac{\pi\alpha^2}{12}\cdot\frac{\F_{2K^0}(s)}{s^{5/2}}\cdot
| A_{bkg} + A_{\rho} + A_{\omega} + A_{\phi} + A_{\omega '} + A_{\rho '}
+ A_{\omega ''} + A_{\phi '} + A_{\rho ''}|^2,
\end{eqnarray*}
where s is the center of mass energy squared, 
$\F_{2K^0}(s) = (s/4 - m^2_{K^0})^{3/2}$ is a phase space factor for two kaons,
$A_{bkg}$ is a constant amplitude of nonresonant background. 
$A_{V} = ( g_{V\gamma}g_{VK\bar K} )\cdot\Delta_V $ are  vector meson
amplitudes and $\Delta_V = (s - m_{V}^2 +\imath\sqrt{s}
\Gamma_{V} (s) )^{-1}$  are vector meson propagators with the energy 
dependence of the width as in \cite{Width}:
\begin{eqnarray*}
  \Gamma_{\rho}(s) & = & \Gamma_{\rho}\cdot\frac{m_{\rho}^2
  \F_{2\pi}(s)} {s \F_{2\pi}(m_{\rho}^2)}, \nonumber \\
 \Gamma_{\omega}(s) & = &  \Gamma_{\omega}\left(B_{2\pi}\frac{m^2_{\omega}
\F_{2\pi}(s)}{s \F_{2\pi}(m^2_{\omega})} + B_{\pi^0\gamma}
\frac{\F_{\pi^0\gamma}(s)}{\F_{\pi^0\gamma}(m^2_{\omega})} + B_{3\pi}
\frac{\sqrt{s} \F_{3\pi}(s)}{m_{\omega} \F_{3\pi}(m^2_{\omega})}\right),
 \nonumber \\
\Gamma_{\phi}(s) & = &
  \Gamma_{\phi}\left(B_{K^+K^-}\frac{m^2_{\phi}\F_{K^+K^-}(s)}
{s \F_{K^+K^-}(m^2_{\phi})} + B_{K_L K_S}\frac{m^2_{\phi}\F_{K_S K_L}(s)}
{s \F_{K_L K_S}(m^2_{\phi})} + \right . \nonumber \\
      & & \left .
+ B_{\eta\gamma}\frac{\F_{\eta\gamma}(s)}
{\F_{\eta\gamma}(m^2_{\phi})} +
  B_{3\pi}\frac{\sqrt{s}\F_{3\pi}(s)}{m_{\phi}
\F_{3\pi}(m^2_{\phi})}\right) \nonumber,  
\end{eqnarray*}
where $B_{f}$ are the branching ratios of the major decay modes of
$\omega$ and $\phi$,  and $\F_f$(s) are corresponding phase space factors. 
For the radiative decay into $P \gamma$ where $P = \eta (\pi^0)$, 
$\F_{P\gamma}(s) = (\sqrt{s}(1-m^2_{P}/s)/2)^3$.  
%$\pi^+ \pi^- \pi^0$ decay mode,   
For the $\F_{3\pi}(s)$ calculation the model assuming the decay 
$V \to \rho \pi \to 3\pi$ was used \cite{kursil}.
The masses and widths of the higher resonances
%$\rho '$ $\omega '$, $\phi '$ 
were taken from \cite{pdg}.
\par The constants $g_{V\gamma}g_{VK\bar K}$ were calculated using
the experimental values of the electronic widths \cite{pdg} and  assuming
SU(3) relations with ideal mixing. The coupling constants of the higher
resonances were parameters of the fit.
Relative phases
between $\rho$ and $\rho '(\rho '')$, $\omega$ and $\omega '(\omega '')$ 
as well as between $\phi$ and $\phi '$
were taken equal to $\pi$ according to \cite{ND}.
\par The cross section can be rewritten in terms of $\sigma_0$ - the
cross section in the resonance peak for the 
$\phi\rightarrow K^0_S K^0_L$ decay mode:
\begin{eqnarray*}
\sigma_K (s) & = & \sigma_0\cdot\frac{\F_{2K}(s)}{\F_{2K}(m^2_{\phi})}\cdot
\frac{m_{\phi}^7\Gamma_{\phi}^2}{s^{5/2}}|-\Delta\rho +
\frac{1}{3}
\Delta\omega + \Delta\phi + \nonumber \\
& &
\frac{g_{\rho '\gamma}g_{\rho ' K\bar K}}{g_{\phi\gamma}g_{\phi K\bar K}}
\Delta\rho ' + \frac{g_{\rho ''\gamma}g_{\rho ''K \bar K}}
{g_{\phi\gamma}g_{\phi K\bar K}} \Delta\rho '' \nonumber \\
&&
-\frac{g_{\omega '\gamma}g_{\omega ' K\bar K}}{g_{\phi\gamma}g_{\phi K\bar K}}
\Delta\omega ' - \frac{g_{\omega ''\gamma}g_{\omega ''K \bar K}}
{g_{\phi\gamma}g_{\phi K\bar K}} \Delta\omega '' \nonumber \\
& &
-\frac{g_{\phi '\gamma}g_{\phi ' K\bar K}}
{g_{\phi\gamma}g_{\phi K\bar K}}\Delta\phi ' + 
\frac{A_{bkg}}{g_{\phi\gamma}g_{\phi K\bar K}}|^2 . \nonumber
\end{eqnarray*}
The non-resonant background amplitude can be written as:
$$
    A_{bkg} = \sqrt{\frac{\sigma_{bkg}}{\sigma_0}}
\frac{1}{m_{\phi}{\Gamma_{\phi}}}\cdot g_{\phi\gamma}g_{\phi K\bar K},
$$
where $\sigma_{bkg}$ is the cross section of the non-resonant background.
\par The values of $\sigma_{bkg}$ as well as coupling constants of the higher
resonances obtained from the fit are consistent with zero and
were fixed at zero to determine the $\phi$ meson parameters.  
%\begin{figure}
%\begin{tabular}{cc}
%\subfigure [] %[$\phi$ meson excitation curve in $\phi\to K_L K_S$ channel
% ( depolarization data )] 
%{\epsfig{file=phi_9496.eps,width=0.49\textwidth}
%\label{fig:phi.a}
%}
%&
%\subfigure []  %[$\phi$ meson excitation curve in $\phi\to K_L K_S$ channel
% ( 1996 data )] 
%{\epsfig{file=delta.eps,width=0.49\textwidth}
%\label{fig:phi.b}
%}
%\\
%\end{tabular} 
%\caption{a - Experimental cross section and $\phi$ meson excitation
%curve in the channel $e^+ e^- \to \phi \to K^0_L K^0_S$ (all data ).
%b - Residuals between the measured cross section and theoretical one
%versus center of mass energy in the each scan.}
%\caption{Experimental cross section and $\phi$ meson excitation curves
%in the channel $e^+ e^-\to\phi\to K^0_L K^0_S$. a - data with depolarization
%energy measurement. b - scans 2 -- 4. }
%\label{fig:phi} 
%\end{figure}
\begin{figure*}
\includegraphics[width=0.48\textwidth]{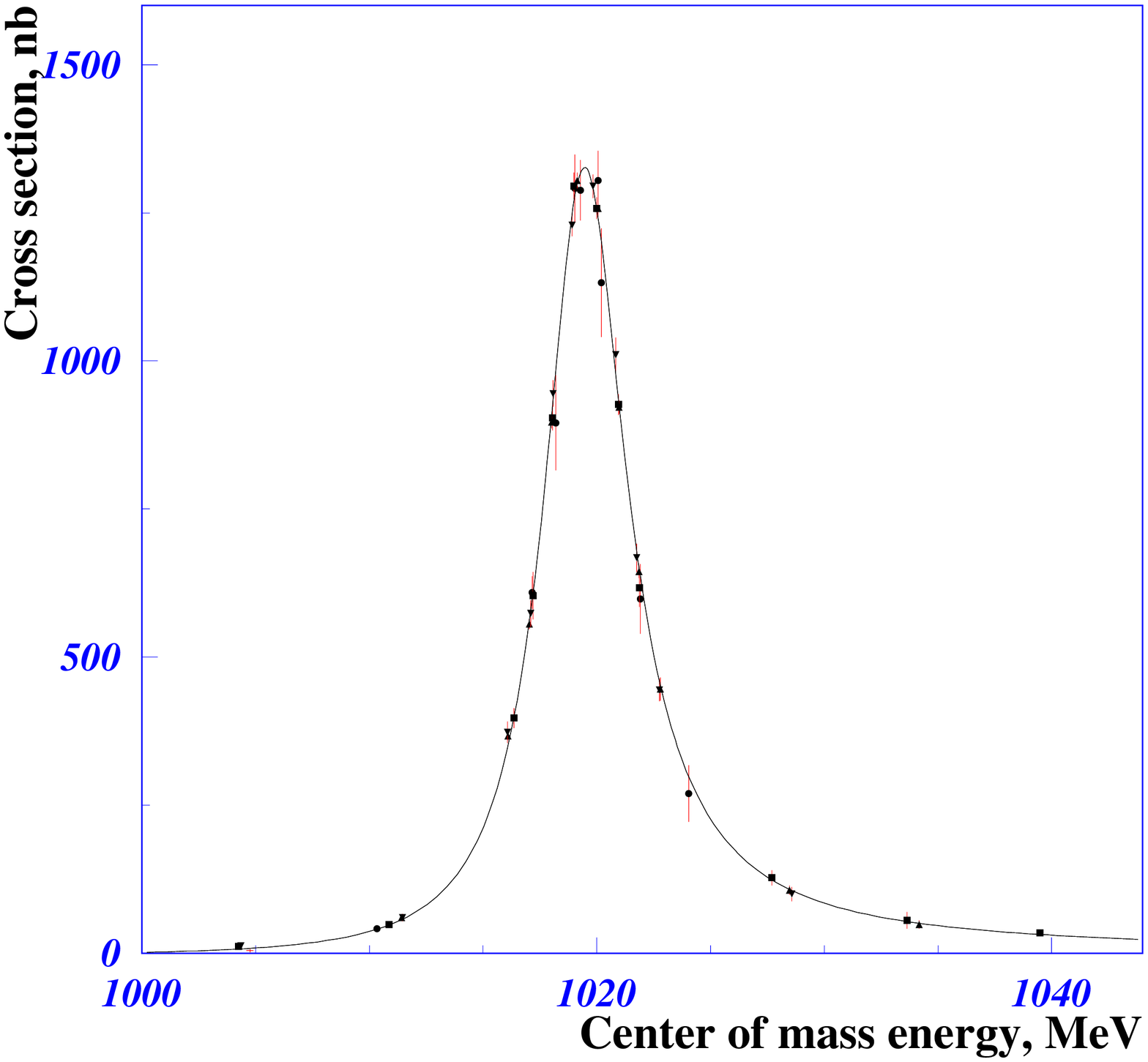}
\hfill
\includegraphics[width=0.48\textwidth]{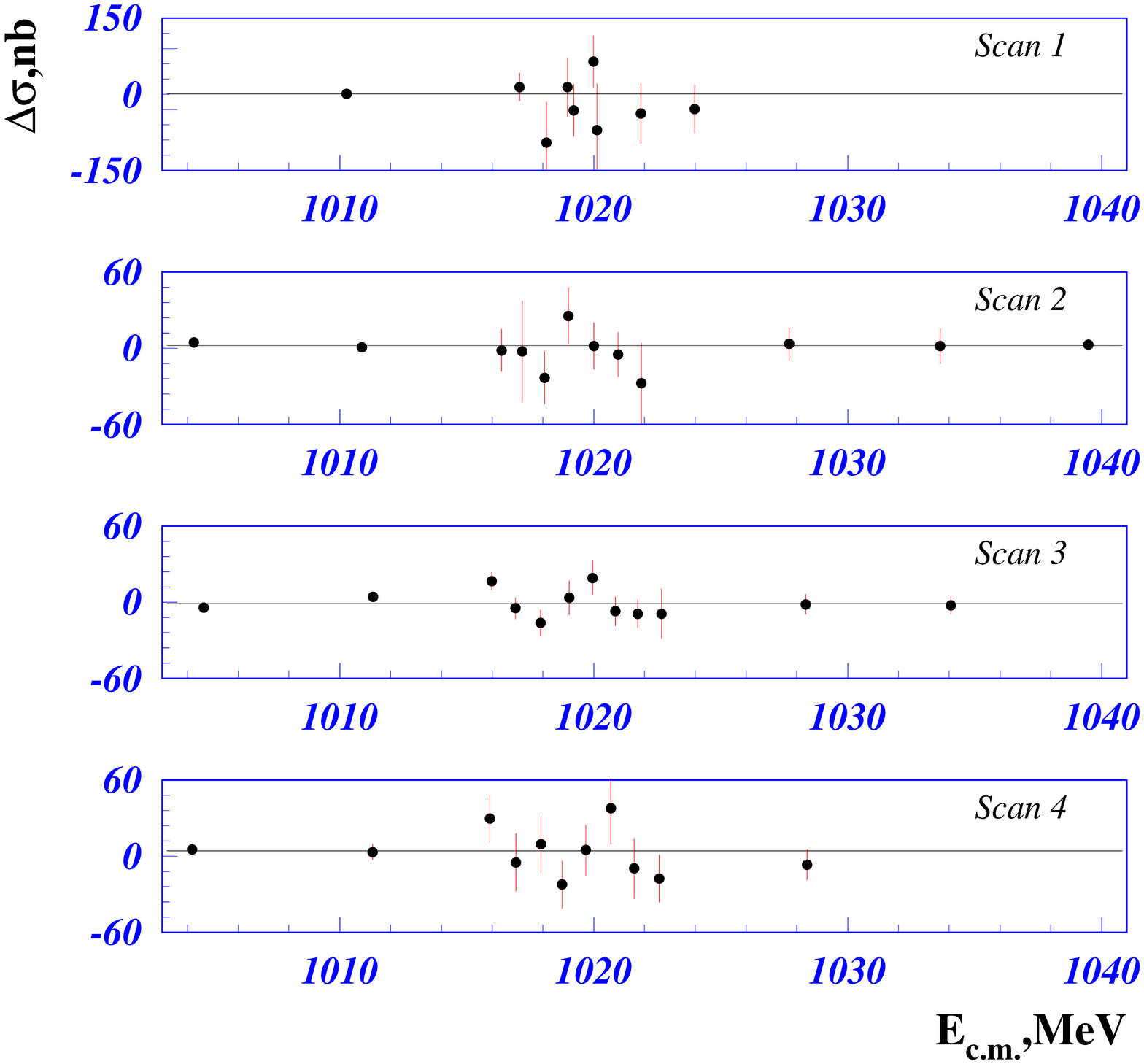}
\\
\parbox[t]{0.48\textwidth}{\caption{Experimental cross section and
$\phi$ meson excitation curve in the channel 
$e^+ e^- \to \phi \to K^0_L K^0_S$ (all data ).}\label{fig:phi}}
\hfill
\parbox[t]{0.48\textwidth}{\caption{Residuals between the measured 
and fitted cross sections versus center of mass energy in each 
scan }\label{delta}}
\end{figure*}
The determination of the beam energy \cite{Prep99} showed that the
energy scale of VEPP-2M had a slope of about
$$
    \frac{\partial E_{scale}}{\partial E_{beam}} \sim 0.005,
$$
where $E_{scale}$ is the beam energy determined and 
$E_{beam}$ is the real beam energy of the collider measured by the 
depolarization method. This slope results in a systematic uncertainty of the 
$\phi$ meson width 
$\frac{\partial E_{scale}}{\partial E_{beam}}\cdot\Gamma_{\phi}\sim$ 0.022 MeV
which is smaller than the statistical error of the width value.
%The systematic uncertainty in the $\phi$ meson width value arised from
%the beam energy determination procedure was estimated to be about
%$\sim$ 20 keV, less than statistical one. 
It allows to consider the width
as a common parameter for all data fits, while the masses and peak
cross sections were different for each scan. As it was found from the
fit:
$$
	\Gamma_{\phi} = 4.477 \pm 0.036 \pm 0.022 \mbox{ MeV},
$$
where the first error is statistical and the second is systematic.
The values of the $\phi$ meson mass and peak cross section obtained
from the fit are presented in Table \ref{phi:par}. 
%The peak cross
%sections and the messes were averaged. The inversed squared
%experimental (both statistical and systematic) errors were used as
%a weights in the averaging procedure. 
The experimental data together with the fitting curve are presented in
Fig.~\ref{fig:phi} where the experimental points of individual scans were 
shifted in accordance with the difference between the average values
of the mass and cross section in the peak and their values in individual
scans.  The residuals between the measured and fitted cross section 
versus center of mass energy are shown in Fig.~\ref{delta} 
for each scan.
%=============================================================
%The experimental data together with the fitting curves are
%presented in Fig. \ref{fig:phi}a,b.
%The $\phi$ meson parameters 
%obtained from the fit for each scan are presented in Table
%\ref{phi:par}, where the first error is statistical and the second
%is systematic.
%==============================================================
%\begin{center}
\begin{table*}
\caption{$\phi$ meson parameters obtained in the experiment.}
\vspace*{5mm}
\begin{tabular}{|c|c|c|c|}
\hline
Scan & $\sigma_0$, nb     & $m_{\phi}$, MeV/c$^2$         & 
$\Gamma_{e^+e^-}$B$\times 10^4$, MeV \\
\hline
 1   & $1370\pm 26\pm 33$ & $1019.487\pm 0.064\pm 0.020$  &
$4.374\pm 0.086\pm 0.105$ \\
 2   & $1332\pm 12\pm 32$ & $1019.558\pm 0.027\pm 0.045$  &
$4.253\pm 0.041\pm 0.102$ \\
 3   & $1374\pm 9\pm 33$  & $1019.444\pm 0.014\pm 0.080$  &
$4.386\pm 0.027\pm 0.105$ \\
 4   & $1396\pm 12\pm 34$ & $1019.399\pm 0.027\pm 0.122$  & 
$4.454\pm 0.043\pm 0.107$ \\
\hline
Average &$1367\pm 15\pm 21$ & $1019.504\pm 0.011\pm 0.033$ & 
$4.364\pm 0.048 \pm 0.065$ \\
\hline
\end{tabular}
% \caption{$\phi$ meson parameters obtained in the experiment.} 
\label{phi:par}
\end{table*}
%\begin{table}[h]
%\begin{tabular}{c}
%\hline
%1 scan \\
%\hline
%$\sigma_{0} = 1306 \pm 28 \pm  33 $ nb \\
%$m_{\phi} = 1019.47\pm 0.06 \pm 0.02 $ MeV/c$^2$ \\
%$\Gamma_{\phi}$ fixed at the value from \cite{pdg} \\
%\hline
%2 scan \\
%\hline
%$\sigma_{0} = 1277 \pm 15 \pm  32 $ nb \\
%$m_{\phi} = 1019.535\pm 0.031 \pm 0.044 $ MeV/c$^2$ \\
%$\Gamma_{\phi} = 4.46 \pm 0.11 \pm 0.03$ MeV  \\
%\hline
%3 scan \\
%\hline
%$\sigma_{0} = 1313 \pm 9 \pm  33 $ nb \\
%$m_{\phi} = 1019.405\pm 0.017 \pm 0.097 $ MeV/c$^2$ \\
%$\Gamma_{\phi} = 4.52 \pm 0.05 \pm 0.03$ MeV  \\
%\hline
%4 scan \\
%\hline
%$\sigma_{0} = 1343 \pm 15 \pm  34 $ nb \\
%$m_{\phi} = 1019.376\pm 0.027 \pm 0.123 $ MeV/c$^2$ \\
%$\Gamma_{\phi} = 4.51 \pm 0.09 \pm 0.05$ MeV  \\
%\hline
%\end{tabular}
%\caption{$\phi$ meson parameters, obtained in the discussed experiment}
%\label{phi:par}
%\end{table}
%\end{center}
\par The systematic
error of the mass from scan 1 in which
the beam energy was measured by the resonance depolarization method 
is determined by  the precision of this technique.
Systematic errors for the mass 
%  and width for
for scans 2 -- 4 were estimated from the difference between energy values 
obtained by two methods of the beam energy determination mentioned above. 
For the second scan it was 0.045 MeV and 
increased to 0.080 MeV and 0.122 MeV for scans 3 and 4 respectively.
% come from the beam energy determination. 
%\par The main contribution to the  systematic error of $\sigma_0$ comes
The systematic error of $\sigma_0$ consists of two parts. The first one
is common for all four scans  and comes 
from the uncertainty of the radiative corrections to the cross section 
of Bhabha events used for the luminosity determination ($\sim$ 1.5\%).
The second part is about 1.9\% and comes from various factors: \\[3mm]  
%Other factors contribute about 1.9\% :\\[3mm]
\noindent
background subtraction \hfill -- 0.3\% ;  \\
reconstruction efficiency \hfill -- 0.7\% ;   \\
trigger efficiency \hfill -- 0.5\% ;\\
radiative corrections for the $e^+ e^- \to K^0_L K^0_S$ process 
\hfill -- 0.5\% ;   \\
correction for decays in flight and nuclear interactions
\hfill -- 0.1\% ;\\
solid angle uncertainty \hfill --  0.9\% ;  \\
uncertainty in beam energy spread \hfill -- 0.2\% ; \\
selection criteria and efficiencies for Bhabha events \hfill -- 1.3\% . \\
\par This part of the systematic error of $\sigma_0$ is numerically the same
(1.9\%) for all scans, although it was independently 
estimated for each scan. Results for the $\phi$ meson mass from
different scans are also independent. Therefore we can average results
from different scans both for the cross section in the peak and mass.
To average data, we used a standard weighted least-squares procedure.
For calculations of the weights statistical and systematic errors 
were added in quadrature. In case of $\sigma_0$ only the variable 
part of the systematic error equal to 1.9\% was added and the 
resulting error was combined with the common error of 1.5\%.   
The average values of the $\phi$ meson parameters are shown in 
the last line of Table 
%\par Since the data in different scans are independent, the
%corresponding $\phi$ meson parameters can be averaged (see Table
\ref{phi:par}.
\par The value of the $\phi$ meson mass obtained in this  
work is the most precise measurement of this parameter among $e^+ e^-$ 
experiments and is 
consistent with all previous results. There is only one
fixed target experiment which gave a more precise value 
of the $\phi$ mass \cite{ZPHY}, differing from ours  
by 2.6 standard deviations.
\par The value of the $\phi$ meson width obtained in our work is 
consistent with all previous results and is the most precise. Note that
its precision surpasses the world average quoted by \cite{pdg}, which is
based on the results of nine previous measurements. 
      
%$\sigma_0 = 1311 \pm 7 \pm 33 $ nb,\\
%$m_{\phi} = 1019.470 \pm 0.013 \pm 0.018$ MeV/c$^2$,\\
%$\Gamma_{\phi} = 4.51 \pm 0.04 \pm 0.02$ MeV.
\par The measurement of $\sigma_0(\phi\to K^0_S K^0_L)$ has been
performed by a number of groups with the results 
presented in Table \ref{sigma0}.
\begin{table*}
\caption{Results of $\sigma_0$ measurements by various groups.}
\vspace*{5mm}
\begin{tabular}{|c|c|c|}
\hline
Group  & $\sigma_0(e^+ e^- \to\phi\to K_L^0 K_S^0)$, nb & Reference \\
\hline 
OLYA,1978 & 1400 $ \pm $ 90                   & \cite{OLYA78} \\
OLYA,1984 & 1440 $ \pm $ 60                   & \cite{OLYA84} \\
ND,  1991 & 1430 $ \pm $143                   & \cite{ND91} \\
SND, 1997 & 1360 $ \pm $ 46                   &  \cite{SND97} \\
\hline
CMD-2,1994,1996 & 1367 $\pm $ 26              & This work   \\
\hline
%\sigma_0(\phi\to K_L^0 K_S^0) & = & (1400 \pm 90) \mbox{
%nb}\cite{OLYA78}\nonumber\\
%\sigma_0(\phi\to K_L^0 K_S^0) & = & (1440 \pm 60) \mbox{
%nb}\cite{OLYA84}\nonumber\\
%\sigma_0(\phi\to K_L^0 K_S^0) & = & (1430 \pm 143) \mbox{
%nb}\cite{ND91}\nonumber\\
%\sigma_0(\phi\to K_L^0 K_S^0) & = & (1360 \pm 46) \mbox{
%nb}\cite{SND97}\nonumber
\end{tabular}
%\caption{Results of $\sigma_0$ measurements 
%$\sigma_0(e^+ e^-\to\phi\to K_L^0 K_S^0)$ by
%by various groups.} 
\label{sigma0}
\end{table*}
\par One can see that our value 
$\sigma_0(\phi\to K^0_L K^0_S) = (1367\pm 26)$ nb 
does not contradict these measurements and is the most precise.
% obtained with $e^+ e^-$ colliding beams.
\par The cross section in the peak obtained in our experiment is
related to the  product
$\Gamma_{e^+e^-}\cdot$B$(\phi\to K^0_L K^0_S)$. To obtain this
value, special fits with this product as a free parameter  
have been performed separately for each scan 
(see the last column of Table~\ref{phi:par}). The averaging procedure
similar to that for $\sigma_0$ gives the following result: 
\begin{eqnarray*}
   \Gamma_{e^+e^-}\cdot\mbox{ B}(\phi\to K^0_L K^0_S) = 
	(4.364\pm 0.048\pm 0.065)\cdot 10^{-4} \mbox{ MeV}.
\end{eqnarray*}
\par
% As it can be seen our value of 
The obtained value of
$\Gamma_{e^+e^-}\cdot B(\phi\to K^0_L K^0_S)$ is the most precise direct
measurement of this product. 

Using $\Gamma_{e^+ e^-}$ from other
experiments, one can obtain $B(\phi\to K^0_L K^0_S)$. For example,
for $\Gamma_{e^+ e^-} = (1.32 \pm 0.04) $ keV from \cite{pdg}, 
$B(\phi\to K^0_L K^0_S) = 0.329 \pm 0.006 \pm 0.010$ has been obtained.

Alternatively,
taking $B(\phi\to K^0_L K^0_S)$ from other works, $\Gamma_{e^+ e^-}$
can be calculated. For $B(\phi\to K^0_L K^0_S) = 0.331 \pm 0.009$ 
from \cite{pdg}, we obtain for the leptonic width 
$\Gamma_{\phi\to ee} =(1.32 \pm 0.02 \pm 0.04) $ keV.   
%This value can be used to calculate either $\Gamma_{e^+e^-}$,
%using $B(\phi\to K^0_L K^0_S)$ from \cite{pdg} or  $B(\phi\to K^0_L K^0_S)$,
%using $\Gamma_{\phi\to ee}$ from \cite{pdg}.
%\par For example, the values $\Gamma_{e^+e^-} = (1.23 \pm 0.01 \pm
%0.06) $ keV or alternatively  
%$B(\phi\to K^0_L K^0_S)  = (0.317 \pm 0.002 \pm 0.020) $
%have been obtained, where 
The first number is  our experimental error (both statistical 
and systematic), while the  second one describes the 
uncertainty of the values from \cite{pdg}.
%\par using  $\gamma_{\phi\to ee}$ from \cite{pdg} with the above
%formula, the following value has been obtained
%$br(\phi\to k^0_l k^0_s)  = 0.317 \pm 0.002 \pm 0.020 $.
% \par values of $\gamma_{\phi\to ee}$ and $br(\phi\to k^0_l k^0_s)$ as well
% as mass and width of $\phi$ obtained in  our experiment are in good
% agreement with the world average from \cite{pdg}. 
\section{Conclusion}
\par Using the CMD-2 data sample of 2.97$\times$10$^5$ $\phi\to K^0_L K^0_S$
events with reconstructed
$K^0_S\rightarrow \pi^+\pi^-$ decay,  
the following values of the $\phi$ meson parameters have been obtained:
\begin{eqnarray*}
         \sigma_{0} & = &   (1367   \pm  15 \pm 21) \mbox{ nb},\nonumber \\ 
 m_{\phi} & = & (1019.504 \pm 0.011 \pm 0.033)\mbox{ MeV/c}^2,\nonumber \\
 \Gamma_{\phi} & = &  (4.477\pm 0.036 \pm 0.022) \mbox{ MeV },\nonumber \\
\Gamma_{e^+e^-}\cdot B(\phi\rightarrow K^0_L K^0_S)
	 & = & ( 4.364 \pm 0.048 \pm 0.065
	)\cdot 10^{-4} \mbox{ MeV} \nonumber.
\end{eqnarray*}
\par These results are more precise than the corresponding 
measurements from any other 
$e^+ e^-$ experiment, and the value of the width is
more precise than all previous measurements. 
% and are in agreement with the  values from
%\cite{pdg}.
\section{Acknowledgements}
\par The authors are grateful to the staff of VEPP-2M for excellent
performance of the collider, to all engineers and technicians who 
participated in the design, commissioning and operation of CMD-2.
\par Special thanks are due to V.N. Ivanchenko for numerous useful
discussions.
 
\end{document}

\maketitle
\hspace*{\parindent} In the analysis described in our Letter a wrong 
sign of the cross section correction for the beam energy spread was chosen.
Although this error only slightly affected the values of the $\phi$ meson
mass and the product of its branching ratios, it unfortunately resulted
in a strong overestimation of the $\phi$ meson total width.
By the time the error was discovered, various improvements in 
the reconstruction and calibration procedures allowed an increase in 
the accuracy of the cross section measurement. 
%A new method of
%calculating radiative corrections for Bhabha events with an accuracy
%about 0.3\% has been developed. 
The resulting systematic uncertainty 
of the cross section and branching ratio measurements is % consists of two 
%parts: 
1.7\% common for all energy scans.\\ % and 1.7\%
%which is scan-to-scan independent. \\

The revised results of our analysis are presented in Table \ref{tab:1}.
Within the errors they are consistent with those in the Letter.
The new corrected value of the $\phi$ meson total width is:
$$
	\Gamma_{\phi} = 4.280 \pm 0.033 \pm 0.025\ MeV.
$$
\begin{table}[h]
\begin{tabular}{|c|c|c|c|}
\hline
Scan & $\sigma_0$, nb     & $m_{\phi}$, MeV/c$^2$ & 
$\Bee\BR\times 10^4$  \\
\hline
 1   & $1365\pm 14\pm 23$ & $1019.506\pm 0.030\pm 0.020$ & 
$0.967\pm 0.010\pm 0.016$\\
 2   & $1342\pm 13\pm 23$ & $1019.512\pm 0.023\pm 0.045$ &
$0.950\pm 0.009\pm 0.016$\\
 3   & $1396\pm 12\pm 24$  & $1019.363\pm 0.017\pm 0.080$ & 
$0.989\pm 0.008\pm 0.017$\\
 4   & $1393\pm 12\pm 24$ & $1019.316\pm 0.021\pm 0.122$ &
$0.986\pm 0.008\pm 0.017$\\
\hline
Average &$1376\pm 6\pm 23$ & $1019.483\pm 0.011\pm 0.025$ &
$0.975\pm 0.004\pm 0.017$\\
\hline
\end{tabular}
\caption{ $\phi$ meson parameters obtained in this analysis}
\label{tab:1}
\end{table}
%\begin{table}[h]
%\begin{tabular}{|c|c|c|c|}
%\hline
%Scan & $\sigma_0
%$, nb      & $m_{\phi}$, MeV/c$^2$ & 
%$\Bee\BR, 10^{-4}$  \\
%\hline
% 1   & $1365\pm 14\pm 24$ & $1019.506\pm 0.030\pm 0.020$ & 
%$0.967\pm 0.010\pm 0.017$  \\
% 2   & $1342\pm 13\pm 23$ & $1019.512\pm 0.023\pm 0.045$ &
%$0.950\pm 0.009\pm 0.017$ \\
% 3   & $1396\pm 12\pm 24$  & $1019.363\pm 0.017\pm 0.080$ & 
%$0.988\pm 0.008\pm 0.017$\\
% 4   & $1392\pm 12\pm 24$ & $1019.316\pm 0.021\pm 0.122$ &
%$0.985\pm 0.009\pm 0.017$ \\
%\hline
%Average &$1374\pm 13\pm 4$ & $1019.483\pm 0.011\pm 0.025$ &
%$0.972\pm 0.009\pm 0.003$ \\
%\hline
%\end{tabular}
%\caption{ $\phi$ meson parameters obtained in this analysis}
%\label{tab:1}
%\end{table}

\par The authors would like to thank their colleagues from the SND 
Collaboration for pointing out this error.

%%\begin{thebibliography}{99}
%\bibitem{phi99}
%{\em R.R.Akhmetshin, E.V. Anashkin, M. Arpagaus, et al.,\/}
%Phys. Lett. {\bf B} 466 (1999), 385.
%\end{thebibliography}